\documentclass[aps,prl,twocolumn,showpacs,floatfix]{revtex4}
\usepackage{graphicx}
\usepackage{dcolumn}
\usepackage{bm}
\usepackage{epsf}
\usepackage{subfigure}
\usepackage{epstopdf}
\usepackage{amsmath}
\usepackage{amssymb}

\newcommand{\bra}[1]{\langle #1|}
\newcommand{\ket}[1]{|#1\rangle}

\newcommand{\mean}[1]{\langle #1 \rangle}
\newcommand{\trace}{{\rm Tr}}

\newcommand{\e}{{\rm e}}

\begin{document}

\title{Entropy production as correlation between system and reservoir}

\author{Massimiliano Esposito}
\altaffiliation[]{Also at Center for Nonlinear Phenomena and Complex Systems,
Universit\'e Libre de Bruxelles, Code Postal 231, Campus Plaine, B-1050 Brussels, Belgium.\\}
\author{Katja Lindenberg}
\affiliation{Department of Chemistry and Biochemistry and Institute for Nonlinear Science, 
University of California, San Diego, La Jolla, CA 92093-0340, USA}
\author{C. Van den Broeck}
\affiliation{Dept WNI, Hasselt University, B 3590 Diepenbeek, Belgium}

\date{\today}

\begin{abstract}
We derive an exact (classical and quantum) expression for the entropy production
of a finite system placed in contact with one or several finite reservoirs each
of which is initially described by a canonical equilibrium distribution. Whereas
the total entropy of system plus reservoirs is conserved, we show that the
system entropy production is always positive and is a direct measure of the
system-reservoir correlations and/or entanglements. Using an exactly solvable
quantum model, we illustrate our novel interpretation of the Second Law in a
microscopically reversible finite-size setting, with strong coupling between
system and reservoirs. With this model, we also explicitly show
the approach of our exact formulation to the standard description of
irreversibility in the limit of a large reservoir.
\end{abstract}
\pacs{05.70.Ln,05.30.-d,05.20.-y}
\maketitle

Starting with the groundbreaking work of Boltzmann, there have been numerous
attempts to construct a microscopic derivation of the Second Law. The main
difficulty is that the prime microscopic candidate for the entropy, namely, the
von Neumann entropy $S=-\trace \rho \ln \rho $ with $\rho$ the density matrix
of the total or compound system, is a constant in time by virtue of Liouville's
theorem.
Related difficulties are the time-reversibility of the microscopic laws and the
recurrences of the micro-states. A common way to bypass these difficulties is
to introduce irreversibility in an ad hoc way, for example by reasoning that the
system is in contact with idealized infinitely large heat reservoirs.
Nevertheless, as was realized early on by Onsager, a consistent description of
the resulting irreversible behavior still carries the undiluted imprint of the
underlying time-reversibility and Liouville's theorem for the system.
Examples are the symmetry of the Onsager coefficients and the
fluctuation dissipation theorem. As examples of more recent discussions we cite
results on work
theorems and fluctuation theorems~\cite{Gallavotti95,Jarzynski97,Seifert05}.
Even more relevant to the question pursued here, we cite the microscopic
expression for the entropy production as the breaking, in a statistical sense,
of the arrow of
time~\cite{VandenBroeck07,VandenBroeck08,Gaspard04b,Gaspard07exp,EspositoReview,
Parrondo2009,Jarzynski06}. We also mention that significant effort has
been devoted to a detailed description and understanding of the interaction
with the heat reservoirs, in particular the difficulties of dealing with the
case of strong coupling~\cite{JarzynskiReply04,09HanggiPRL}.

In this letter we show that the problem of entropy production can be addressed
within a microscopically exact description of a finite system, without
resorting to infinitely large heat reservoirs and without any assumption of weak
coupling. Whereas the von Neumann entropy of system plus reservoirs is
conserved, the entropy production of the system is always positive, even though
it
displays oscillations and recurrences typical of the finite total system.
Interestingly, this entropy production is expressed in terms of the correlations
and/or entanglement between system and reservoirs, so that its positivity can be
explained by a corresponding negative entropy contribution contained in the
correlations and/or entanglement with the reservoirs. As the size of the
reservoirs increases, the recurrences die out, the negative entropy contribution
is diluted in an intricate way over the increasing number of correlations with
reservoir degrees of freedom, and the entropy production of the system itself
approaches the standard thermodynamic form. We will illustrate this novel
interpretation of the Second Law on an exactly solvable  model, namely, a spin
interacting with an $N$-level quantum system via a
random matrix coupling. We focus on the derivation for the quantum case, but the
analogous treatment for the classical system is straightforward. 
 
The set-up is as follows. We consider one or several finite quantum systems $r$
which play the role of finite-size heat reservoirs. Accordingly, their density
matrices $\rho_r(t)$ at the initial time $t=0$ are assumed to be of the
canonical equilibrium form,
\begin{eqnarray}
\rho_r(0)=\rho_r^{\rm eq}=\exp{\big(-\beta_r H_r\big)}/Z_r. \label{ICond}
\end{eqnarray}
Here $\beta_r$, $ H_r$ and $Z_r$  are the corresponding inverse temperature at
$t=0$ (Boltzmann's constant $k_B$ is set equal to $1$),  the Hamiltonian, and
the
partition function at $t=0$.  Being reservoir systems, it is further natural to
assume that their Hamiltonians $H_r$ are  time-independent. At time $t=0$ we
connect a finite quantum system $s$, characterized by Hamiltonian $H_s(t)$ and
density matrix $\rho_s(t)$, to the reservoirs by switching on an interaction
Hamiltonian $V(t)$.  The initial state of the compound system,
characterized by  the density matrix  $\rho(t)$, does not display any
entanglement or correlation,
\begin{eqnarray}
\rho(0) = \rho_s(0) \prod_r  \rho_r^{\rm eq} \label{ICondTR}.
\end{eqnarray}
Correlations and/or entanglements do develop in the subsequent time
evolution of $\rho(t)$, which obeys Liouville's equation for the total
Hamiltonian 
\begin{eqnarray}
H(t)=H_s(t)+\sum_r H_r+V(t) . \label{Htot}
\end{eqnarray}
Note that in addition to the issue of relaxation of a system in contact with
a reservoir, this scenario includes the
ingredients for the study a driven system, cf. the time-dependence of
the system's Hamiltonian, as well as that of a nonequilibrium steady state,
which can be realized in view of the presence of several heat reservoirs. In
fact, the above construct can easily be generalized to include particle
reservoirs described via grand-canonical distributions. This would allow
the consideration of particle flows in addition to heat flows.

We are primarily interested in the occurrence and characterization of 
irreversible behavior in the system, and we thus focus our attention on
the entropy $S(t)$ of the system,
\begin{eqnarray}
S(t) \equiv -\trace_s \rho_s(t) \ln \rho_s(t)
\label{vonNeumannEntropy}
\end{eqnarray}
where $\rho_s(t)$ is the trace of $\rho(t)$ over the degree of freedom of all the reservoirs.
Contrary to the total von Neumann entropy, the entropy of the system is in general a function 
of time, technically speaking because the dynamics of $\rho_s(t)$ is not unitary. 
More to the point for the ensuing discussion, we note that from the 
thermodynamic point of view we are dealing with an energetically open system. 
We now show that it is precisely the time invariance of the total von Neumann entropy 
which induces a natural separation of the entropy change of the system into separate
contributions from an entropy flow and an entropy production. Using
$-\trace \rho(t) \ln \rho(t)=-\trace \rho(0) \ln \rho(0)
=-\trace_s \rho_s(0) \ln \rho_s(0)-\sum_r\trace_r \rho_r^{\rm eq} \ln \rho_r^{\rm eq}$,
we find for the entropy change of the system
 \begin{eqnarray}\label{dsc}
&&\hspace{-0.6cm}\Delta S(t)=S(t)-S(0) \nonumber \\
&=&-\trace \rho(t) \ln \rho_s(t)+\trace \rho(t) \ln \rho(t)-\sum_r\trace_r \rho_r^{\rm eq} \ln \rho_r^{\rm eq} \nonumber\\
&=&-\trace \rho(t) \ln\{ \rho_s(t) \prod_r\rho_r^{\rm eq}\}+\trace \rho(t) \ln \rho(t) \nonumber \\
&&+\sum_r\trace_r[\rho_r(t)- \rho_r^{\rm eq}] \ln \rho_r^{\rm eq}. 
\end{eqnarray}
We conclude that the change in the entropy of the system can be written in the
standard thermodynamic form~\cite{Prigogine}
\begin{eqnarray}
\Delta S(t)=\Delta_i S(t)+\Delta_e S(t).
\label{central}
\end{eqnarray}
The entropy flow, representing the reversible contribution to the system entropy
change due to heat exchanges, is identified as the last term in (\ref{dsc}).
After some manipulation using the explicit form of $\rho_r^{\rm eq}$, it can be written as
\begin{eqnarray}
\Delta_e S(t)  = - \sum_r \beta_r(\mean{H_r}_t-\mean{H_r}_0),
\label{EntropyFlow}
\end{eqnarray}
where $\mean{\bullet}_t \equiv \trace[\rho(t)\bullet]$.
Of particular interest is the resulting expression for the entropy production,
\begin{eqnarray}
\Delta_i S(t) \equiv D[\rho(t)||\rho_s(t)\prod_r \rho_r^{\rm eq}],
\label{EntropyProd}
\end{eqnarray}
which represents the irreversible contribution to the entropy
change of the system. Here,  $D[\rho||\rho']$ is the quantum relative entropy
between two density matrices $\rho$ and $\rho'$,
\begin{eqnarray}
D[\rho||\rho'] \equiv \trace \rho \ln \rho - \trace \rho \ln \rho'.
\label{DefRelEntropy}
\end{eqnarray}
It has the following important properties \cite{Nielsen,Breuer02}. 
The relative entropy is positive, and equal to zero only when the two matrices
are identical. We thus conclude that the entropy production introduced above is
indeed a positive quantity, $\Delta_i S(t)\geq 0$, and vanishes only when
the system and the reservoirs are totally decorrelated. Furthermore, the
relative entropy is a measure of the ``distance'' between two density matrices.
Hence, as announced earlier, the entropy production explicitly expresses how
``far'' the actual state $\rho(t)$ of the total system is from the
decorrelated/disentangled product state $\rho_s(t) \prod_r \rho_r^{\rm eq}$. 

To further clarify the significance of our central result (\ref{central}), we
make a number of additional comments.
First, starting with Eq.~(\ref{central}) we can rewrite the entropy production
as $\Delta_i S(t)=\Delta S(t)-\Delta_e S(t)$. $\Delta S(t)$  is the exact
entropy change of the system. If one assumes that the entropy change in each
heat reservoir is given by $\Delta S_r(t)=-\beta_r Q_r(t)$, and if one further
erroneously supposes that the total entropy is simply the sum of the system and 
reservoir entropy, one  concludes that the positive entropy production $\Delta_i S(t)$ 
is the entropy increase in the total system. This is of  course in flagrant
contradiction with the premise that led to the identification of $\Delta_i
S(t)$, namely, that the entropy of the total system remains unchanged.
The error resides in disregarding a contribution $-\Delta_i S(t)$ to the total
entropy, which is precisely the negative entropy contribution contained in the
correlations and entanglement between system and reservoir. The argument may
on the surface appear circular, but the neglect of the negative
entropy contribution is actually quite natural from an operational point of
view: while one has full microscopic access to the system's properties, one 
only  controls or measures the energy and no other properties of the reservoir.
In this sense, the above procedure leading to an apparent total positive entropy
change can be viewed as a coarse graining operation that retains the full
microscopic description of the system but reduces the reservoirs plus
correlations to an idealized heat reservoir description. In the limit
of large reservoirs it is likely that this latter description deviates
very little from the canonical distribution. Concerning the correlations, one
expects that they will be diluted over the exponentially many higher-order
correlations, becoming in effect irretrievable. Furthermore, this will happen
exponentially fast in time if the reservoirs display non-integrable, chaotic 
properties.  

Second, while $\Delta_i S(t)$ is a positive quantity, it does not increase
monotonically in time. In fact, oscillations are bound to arise in view of the
recurrences in the state of the finite total system. In this respect, it is
important to stress that we consider the entropy change starting from the
natural but specific initial condition (\ref{ICondTR}). The transient decreases
of $\Delta_i S(t)$ can be interpreted as the reappearance of the negative
entropy, hidden in the correlations, as system and reservoir transiently return
to states close to this decoupled initial state. In the limit of large
reservoirs, recurrences will become less and less likely, and $\Delta_i S(t)$ is
expected to converge to a convex monotonically increasing function of $t$. 

Third, we make the connection with a recent discussion
\cite{JarzynskiReply04,09HanggiPRL} concerning the appropriate definition of
work and free energy in a driven system strongly coupled to a heat reservoir.
We consider the case of a single reservoir at temperature $T=\beta^{-1}$,
for convenience dropping the subscript $r$.
Using the fact that $\trace H(t) \dot{\rho}(t)=0$, the work done on the total
system can be written as 
\begin{eqnarray}
W &\equiv&
\mean{H(t)}_t-\mean{H(0)}_0 \nonumber\\
&=& \int_{0}^{t}d\tau
\trace \big( \dot{H_s}(t)+\dot{V}(t) \big) \rho(t).
\label{work}
\end{eqnarray}
The change of the energy of the system, including the contribution of
the interaction term, reads
\begin{eqnarray}
\Delta U(t) \equiv \mean{\big( H_s(t)+ V(t) \big)}_t-\mean{\big( H_s(0)+ V(0) \big)}_0 .
\label{SysEnergy}
\end{eqnarray}
Using $\trace \big( H_s(t)+ V(t) \big) \dot{\rho}(t) = -\trace H_r
\dot{\rho}(t)$, we find that this energy change can be written, in accordance
with the First Law, as the sum of work and heat,
\begin{eqnarray}
\Delta U(t) = W(t) + Q(t).
\label{FirstPrinc}
\end{eqnarray}
Next, introducing the nonequilibrium free energy
\begin{eqnarray}
\Delta F(t) \equiv \Delta U(t) - T \Delta S(t), \label{FreeEnergy}
\end{eqnarray}
we can rewrite the expression (\ref{central}) for the entropy production
in the
standard thermodynamic form for a driven system in contact with a heat
reservoir,
\begin{eqnarray}
T \Delta_i S(t) = W(t)-\Delta F(t) \geq 0 .  \label{EntropyProdAlter}
\end{eqnarray}
This expression is exact. If we assume that the total system relaxes to a final
canonical equilibrium at temperature $\beta^{-1}$, this nonequilibrium free
energy difference reduces to the equilibrium  expression identified in the
context of the work theorem in both the weak coupling
\cite{TalknerCampisi08,VandenBroeck07,Jarzynski97} and strong coupling regimes 
\cite{JarzynskiReply04,09HanggiPRL}.

Finally, we discuss the connection with the following alternative definition for the irreversible entropy change, proposed in open quantum system 
theory \cite{Breuer02}:
\begin{eqnarray}
\Delta_i \bar{S}(t) &\equiv& D[\rho_s(0)||\rho_s^{\rm eq}]-D[\rho_s(t)||\rho_s^{\rm eq}] \nonumber \\
&=&\Delta S(t) - \Delta_e \bar{S}(t) .
\label{EntropyProdBreuer}
\end{eqnarray}
The entropy flow is now defined as $\Delta_e \bar{S}(t) \equiv \beta
(\mean{H_s}_t-\mean{H_s}_0)$ and $\rho_s^{\rm eq}=\exp{(-\beta H_s)}/Z_s$. To
compare this expression with our definition (\ref{EntropyProd}) for the entropy
production, we note that total energy is conserved by the dynamics, $\mean{H}_t
=\mean{H}_0$, and hence
\begin{eqnarray}
\Delta_i S(t)  = \Delta_i \bar{S}(t) - \beta (\mean{V}_t-\mean{V}_0) \geq 0 .
\label{comparison}
\end{eqnarray}
The two definitions thus differ by the interaction term, which vanishes in the
limit of either weak coupling or high temperature. The definition
(\ref{EntropyProdBreuer}) has the obvious advantage of being exclusively
expressed in terms of the system density matrix, while our definition
(\ref{EntropyProd}) requires the total density matrix. However, we will show
that contrary to our expression, ~(\ref{EntropyProdBreuer}) is
not always a positive quantity. The positivity of
(\ref{EntropyProdBreuer}) can be proven when $\rho_s^{\rm eq}$ is the stationary
solution of the reduced dynamics \cite{Breuer02}. This will generically be the
case in the weak-interaction large-reservoir limit, i.e., precisely when
the interaction term in (\ref{comparison}) can be neglected and
(\ref{EntropyProdBreuer}) becomes identical to (\ref{EntropyProd}). 
An even stronger statement can be made when, in the same limit, the system
dynamics can be described by a Markovian quantum master equation of the form
$\dot{\rho}_s(t)={\cal L} \rho_s(t)$, where ${\cal L}$ is a Redfield
superoperator satisfying ${\cal L} \rho_s^{\rm eq}=0$
\cite{Breuer02,GaspNaga99}. Under these conditions, it is known that the entropy
production is a convex functional of the system density matrix \cite{Breuer02},
with a positive rate of entropy production $\frac{d}{dt} \Delta_i S(t) \approx
\frac{d}{dt} \Delta_i \bar{S}(t) \geq 0$.

We will now illustrate the above findings in a two-level quantum spin
coupled to an $N$-level reservoir via a random matrix.
The total Hamiltonian reads
\begin{eqnarray}
H=\frac{\Delta}{2} \sigma_z + H_r + \lambda \sigma_x R .
\label{modelhamiltonian}
\end{eqnarray}
$\sigma_{x,z}$ are the well known Pauli matrices. The reservoir Hamiltonian
$H_r$ is a diagonal matrix with $N$ equally spaced eigenvalues between $-0.5$
and $0.5$. The coupling matrix is $R=X/\sqrt{8N}$, where $X$ is a
Gaussian orthogonal random matrix of size $N$ with probability density
proportional to $\exp(-\frac{1}{4}{\rm Tr}{X}^2)$ \cite{MehtaBook}. This
model is similar to the spin-GORM model of Ref.~\cite{EspoGasp03b}. 
The system is initially assumed to be in the pure lower energy state
$\rho_s(0)=\ket{0}\bra{0}$, where $\sigma_z \ket{0}=- \ket{0}$, and the reservoir
is initially in a canonical equilibrium state at the temperature $\beta^{-1}$.
In the weak-coupling large-reservoir limit, the resulting Redfield equation
leads to the following closed relaxation equation for the z-component of the
spin ($\hbar=1$):
\begin{eqnarray}
\mean{\sigma_{z}}_t&=& \mean{\sigma_{z}}_{\rm eq} + (\mean{\sigma_{z}}_0 - \mean{\sigma_{z}}_{\rm eq}) e^{-\gamma t}\nonumber\\
\gamma &=& 2 \pi \lambda^2 (\tilde{\alpha}(\Delta)+\tilde{\alpha}(-\Delta)) \nonumber\\
\mean{\sigma_{z}}_{\rm eq} &=& \frac{\tilde{\alpha}(-\Delta)-\tilde{\alpha}(\Delta)}
{\tilde{\alpha}(-\Delta)+\tilde{\alpha}(\Delta)}.
\label{solequi}
\end{eqnarray}
Here $\tilde{\alpha}(\omega)$ is the Fourier transform of the reservoir
correlation function $\alpha(t)=\trace_r \rho_r^{\rm eq} \exp{[i H_r t]} R
\exp{[-i H_r t]} R$,
\begin{eqnarray}
\tilde{\alpha}(\vert \omega \vert) = \frac{1}{16} \frac{\e^{-\beta/2} \e^{\beta \vert \omega \vert}-\e^{\beta/2}}
{\e^{-\beta/2} + \e^{\beta/2}}= \e^{\beta \vert \omega \vert}
\tilde{\alpha}(-\vert \omega \vert).
\end{eqnarray}
The $x$ and $y$ components of the spin evolve independently of the $z$
component, and are zero for our initial condition.
\begin{figure}[h]
\centering
\rotatebox{0}{\scalebox{0.35}{\includegraphics{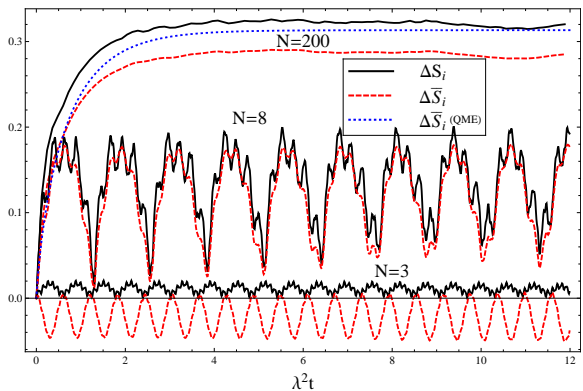}}}
\caption{(Color online)
$\Delta S_i$ [resp. $\Delta \bar{S}_i$] is the entropy production (\ref{EntropyProd}) [(\ref{EntropyProdBreuer})] calculated using the exact numerical dynamics. $\Delta \bar{S}_i$(QME) is the irreversible entropy production (\ref{EntropyProdBreuer}) calculated using the Redfield equation (\ref{solequi}). Parameters are $\Delta=0.1$, $\beta=10$, $\lambda=0.1$.}
\label{plot1}
\end{figure}

We are now in a position to compare the definition
(\ref{EntropyProdBreuer}) of the entropy change with the irreversible
entropy change (\ref{EntropyProd}) that
follows from Redfield theory. This is accomplished through an exact numerical
solution of our model for finite $N$. The results are summarized in Fig.
\ref{plot1}. Note that we show single realizations of the random matrix. For
small $N$, we observe a pronounced oscillatory behavior and even
near-recurrences
very close to zero of our entropy change (\ref{EntropyProd}). While the latter
always remains positive, the entropy change (\ref{EntropyProdBreuer}) can be
negative for small values of $N$, which is clearly not acceptable. In the limit
of a large reservoir ($N \rightarrow \infty$ ), both expressions converge to one
another and coincide with the positive and convex irreversible entropy change
predicted by the Redfield equation.

We conclude that (\ref{EntropyProd}) is a proper definition for the entropy
change, one that remains valid for a small system strongly coupled to small
reservoirs. In the limit of large reservoirs, it converges to a convex
irreversible entropy, coinciding with the familiar definition of entropy
production \cite{Breuer02} for the quantum master equation. Our identification
of the entropy production (\ref{EntropyProd}) within an exact microscopic
framework vindicates the description of irreversibility as a property of
open systems, with the entropy production rather than the entropy of the 
total system playing the central role. The microscopic origin of 
the entropy production, explained in terms of correlations established
between the system and its reservoirs, is reminiscent of Boltzmann's
Stosszahlansatz. However, our analysis of the  micro-dynamics and the
identification of the entropy production of the system are exact. The
appearance of irreversibility  as the omission, in the reservoirs, of its
correlations with the system, provides a natural, precise and transparent
interpretation of the Second Law.

\begin{acknowledgments}
M. E. is supported by the FNRS Belgium (charg\'e de recherches) and 
by the Luxembourgish government (Bourse de formation recherches). K.L. and C.
VdB. gratefully acknowledge the support of the US National Science Foundation
through Grant No. XXXXX.
\end{acknowledgments}

\end{document}